\documentstyle[12pt,epsf,rotate]{article}

\newcommand{\tr}{\mbox{tr}}
\newcommand{\eqn}[1]{(\ref{#1})}

\newcommand{\nn}{\nonumber}

\newcommand{\real}{{\bb R}} %% real numbers
\font\mybb=msbm10 at 12pt
\def\bb#1{\hbox{\mybb#1}}

\def\e{{\rm e}}
\def\beq{\begin{equation}}
\def\eeq{\end{equation}}
\def\be{\begin{equation}}
\def\ee{\end{equation}}
\def\bea{\begin{eqnarray}}
\def\eea{\end{eqnarray}}
\def\bd{\begin{displaymath}}

\makeatletter
\newdimen\normalarrayskip              % skip between lines
\newdimen\minarrayskip                 % minimal skip between lines
\normalarrayskip\baselineskip
\minarrayskip\jot
\newif\ifold             \oldtrue            
\def\arraymode{\ifold\relax\else\displaystyle\fi} % mode of array entries
\def\@arrayskip{\ifold\baselineskip\z@\lineskip\z@
     \else
     \baselineskip\minarrayskip\lineskip2\minarrayskip\fi}
\def\@arrayclassz{\ifcase \@lastchclass \@acolampacol \or
\@ampacol \or \or \or \@addamp \or
   \@acolampacol \or \@firstampfalse \@acol \fi
\edef\@preamble{\@preamble
  \ifcase \@chnum
     \hfil$\relax\arraymode\@sharp$\hfil
     \or $\relax\arraymode\@sharp$\hfil
     \or \hfil$\relax\arraymode\@sharp$\fi}}
\def\@array[#1]#2{\setbox\@arstrutbox=\hbox{\vrule
     height\arraystretch \ht\strutbox
     depth\arraystretch \dp\strutbox
     width\z@}\@mkpream{#2}\edef\@preamble{\halign \noexpand\@halignto
\bgroup \tabskip\z@ \@arstrut \@preamble \tabskip\z@ \cr}%
\let\@startpbox\@@startpbox \let\@endpbox\@@endpbox
  \if #1t\vtop \else \if#1b\vbox \else \vcenter \fi\fi
  \bgroup \let\par\relax
  \let\@sharp##\let\protect\relax
  \@arrayskip\@preamble}
\makeatother

\newlength{\extraspace}
\newlength{\extraspaces}
\begin{document}

\renewcommand{\footnotesize}{\small}

\addtolength{\baselineskip}{.8mm}

\thispagestyle{empty}

\begin{flushright}
\baselineskip=12pt
{\sc OUTP}-98-56P\\
gr-qc/9807070\\
\hfill{  }\\ July 1998
\end{flushright}
\vspace{.5cm}

\begin{center}

\baselineskip=24pt

{\Large\bf{Spacetime Quantization \\from Non-abelian D-particle
Dynamics}}\\[15mm]

\baselineskip=12pt

{\sc Nick E. Mavromatos\footnote{PPARC Advanced Fellow (U.K.).\\ E-mail: {\tt
n.mavromatos1@physics.oxford.ac.uk}}} {\sc
and Richard J.\ Szabo\footnote{Work supported in part by PPARC (U.K.). Address
after September 1, 1998: The Niels Bohr Institute, University of Copenhagen,
Blegdamsvej 17, DK-2100 Copenhagen \O, Denmark.\\ E-mail: {\tt
r.szabo1@physics.oxford.ac.uk}}}
\\[5mm]
{\it Department of Physics -- Theoretical Physics\\ University of Oxford\\ 1
Keble Road, Oxford OX1 3NP, U.K.} \\[15mm]

\vskip 1.0 in

{\sc Abstract}

\begin{center}
\begin{minipage}{15cm}

We describe the short-distance properties of the spacetime of a system of
D-particles by viewing their matrix-valued coordinates as coupling constants of
a deformed worldsheet $\sigma$-model. We show that the Zamolodchikov metric on
the associated moduli space naturally encodes properties of the non-abelian
dynamics, and from this we derive new spacetime uncertainty relations directly
from the quantum string theory. The non-abelian uncertainties exhibit
decoherence effects which suggest the interplay of quantum gravity in multiple
D-particle dynamics.

\end{minipage}
\end{center}

\end{center}

\bigskip

\begin{flushleft}
\end{flushleft}

\vfill
\newpage
\pagestyle{plain}
\setcounter{page}{1}

An old idea in string theory is that the classical ideas of general relativity
at very small distance scales break down. If one uses only string states as
probes of short distance structure, then the usual concept of spacetime ceases
to make sense beyond the intrinsic finite length $\ell_s$ of the string itself
\cite{ven}. In recent years, however, the discovery of certain solitonic
structures, known as D-branes \cite{polchinski}, has drastically changed the
understanding of the nonperturbative and spacetime properties of string theory.
The simplest such string solitons are known as D-particles which are point-like
objects to which the endpoints of open strings can attach. It was shown in
\cite{liyoneya} that, for the non-relativistic scattering of two D-particles
with impact parameter of order $\Delta y_i$, the space-time uncertainty
principle
\beq
\Delta y_i\,\Delta t\geq\ell_s^2
\label{tyuncert}\eeq
yields spatial and temporal indeterminancies $\Delta y_i\geq g_s^{1/3}\ell_s$
and $\Delta t\geq g_s^{-1/3}\ell_s$, where $g_s$ is the string coupling
constant and the former bound coincides with the 11-dimensional Planck length
$\ell_{\rm P}$ which is the characteristic distance scale of M-theory
\cite{witten1}. Thus with weak string interactions, D-particles can probe
distances much smaller than $\ell_s$.

The target space dynamics of a system of $N$ D-particles is usually described
by a matrix quantum mechanics which is obtained by dimensionally reducing
ten-dimensional $U(N)$ Yang-Mills theory to the worldlines of the D-particles
\cite{bound}. The D-particle coordinates are given by $N\times N$ Hermitian
matrices $Y^i_{ab}$ which are to be thought of as adjoint Higgs fields. The
diagonal components $Y^i_{aa}$ represent the positions of the $N$ D-particles
themselves while the off-diagonal components $Y^i_{ab}$ correspond to short
oriented open string excitations between D-particles $a$ and $b$ when the two
objects are brought infinitesimally close to each other (fig. 1). In this
letter we shall adopt a different formalism by treating the $Y_{ab}^i$ as
coupling constants of a deformed worldsheet conformal field theory
\cite{emn,lm}, which is the pertinent field theory for the weak-coupling regime
of the dynamics. We will show that the small-scale structure of spacetime is
naturally encoded within the Zamolodchikov metric \cite{zam} on the moduli
space $\cal M$ of $\sigma$-model couplings, which itself captures all of the
non-trivial dynamical information about the D-particle system. Quantization of
$\cal M$ is achieved by summing over worldsheet genera of the deformed
$\sigma$-model. This promotes the D-particle couplings to quantum operators in
target space from which non-trivial uncertainty relations can be derived. In
this way we will find, directly from the quantum string theory itself, new
quantum mechanical uncertainties which are particular to the non-abelian nature
of D-particle dynamics, in addition to the uncertainties implied by
\eqn{tyuncert}. These smearings occur among the open string excitations between
different D-particles and they yield a triple space-time uncertainty relation
which implies that the scattering of D-particles at high energies can probe
very small distances through their string interactions. We will also show that
there is a smearing with respect to different {\it spatial} string coordinates
which is rather different in spirit than the short-large distance duality
relation \eqn{tyuncert}. This represents a proper quantization of the
noncommutative spacetime implied by D-particle field theory and thereby yields
limitations on the accuracy of simultaneously locating different spatial
directions. These correlations are also given by energy-dependent distances, so
that the non-abelian uncertainties exhibit decoherence effects which are
characteristic of quantum gravity. This construction therefore illuminates the
manner in which D-particle interactions probe very short distances where the
effects of quantum gravity are significant.

\begin{figure}[htb]
\epsfxsize=0.75in
\bigskip
\centerline{\rotate{\rotate{\rotate{\epsffile{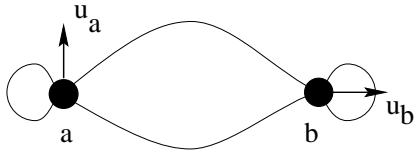}}}}}
\caption{\it\baselineskip=12pt Emergence of $U(N)$ gauge symmetry for bound
states of $N$ D-particles (solid circles). An oriented fundamental string (thin
lines) can start and end either at the same or different D-particle, with
energy proportional to their separation, giving $N^2$ massless vector states
when the particles are practically on top of each other. These states form a
representation of $U(N)$. For the configuration shown, D-particles $a$ and $b$
move at mutually transverse velocities $u_a$ and $u_b$ in target space.}
\bigskip
\label{dbrane}\end{figure}

The tree-level $N\times N$ matrix D-particle dynamics is described by the path
integral \cite{polchinski}
\beq
Z_N[Y]=\left\langle W[\partial\Sigma;Y]\right\rangle_0\equiv\int
Dx~\e^{-S_0[x]}~\tr\,{\cal P}\exp\left(\frac{ig_s}{\ell_s^2}
\oint_{\partial\Sigma}Y_i(x^0(s)){}~dx^i(s)\right)
\label{partfatbrane}\eeq
where $S_0[x]=(1/2\ell_s^2)\int_\Sigma(\partial x^\mu)^2$ is the free
worldsheet $\sigma$-model action which is defined on a disc $\Sigma$ whose
boundary $\partial\Sigma$ is a circle parametrized by $s\in[0,1]$. The
elementary string fields $x^\mu$ are maps from $\Sigma$ into flat
9+1-dimensional spacetime, and the path-ordered Wilson loop operator in
\eqn{partfatbrane} is written in the temporal gauge for the
dimensionally-reduced $SU(N)$ Chan-Paton gauge field $A_i=(1/\ell_s^2)Y_i$. To
write \eqn{partfatbrane} in the form of a local deformation of the conformal
field theory $S_0[x]$, we disentangle the path ordering by introducing
one-dimensional complex auxilliary fields $\bar\xi_a(s),\xi_b(s)$ on
$\partial\Sigma$ \cite{dorn} and writing
\beq
W[\partial\Sigma;Y]=\int D\bar\xi~D\xi~\bar\xi_c(0)
\exp\left[-\int_0^1ds~\bar\xi_a(s)\left(\delta^{ab}\partial_s+
\mbox{$\frac{ig_s}{\ell_s^2}$}\,Y_i^{ab}(x^0(s))\partial_s
x^i(s)\right)\xi_b(s)\right]\xi_c(1)
\label{wilsonloopaux}\eeq
The auxilliary quantum fields have propagator
$\langle\bar\xi_a(s)\xi_b(s')\rangle=\delta_{ab}\Theta(s'-s)$, where
$\Theta(s)$ is the usual step function. The partition function
\eqn{partfatbrane} now takes the form of a path integral involving a local
action $S_0[x]+\oint_{\partial\Sigma}Y_i^{ab}(x^0(s))V_{ab}^i(x;s)$ and
matrix-valued vertex operators
$V_{ab}^i(x;s)=(ig_s/\ell_s^2)\partial_sx^i(s)\bar\xi_a(s)\xi_b(s)$.

The emergence of non-trivial spacetime uncertainty relations can be seen most
directly by treating the D-particles as heavy non-relativistic objects. Their
collective coordinates can then be described by recoil operators \cite{kmw}
which are induced by the scattering of string matter off the D-particle
background. The deformation is given by $Y_i^{ab}(x^0)=
\ell_s(Y_i^{ab}\ell_s\epsilon+U_i^{ab}x^0)\epsilon\Theta_\epsilon(x^0)$, where
the D-particle spatial coordinates are identified with the couplings $Y_i$ and
the $U_i$ correspond to their Galilean recoil velocities. The parameter
$\epsilon\to0^+$ regulates the ambiguous value of $\Theta(s)$ at $s=0$, which
ensures that the D-particle system starts moving only at time $x^0=0$. It is
related to the worldsheet ultraviolet cutoff scale $\Lambda$ (measured in units
of the size of $\Sigma$) by $\epsilon^{-2}=2\ell_s^2\log\Lambda$. For finite
$\epsilon$, these operators each have an anomalous dimension
$-\frac12|\epsilon|^2<0$ \cite{kmw} and thus lead to a relevant deformation of
$S_0[x]$. The corresponding $\beta$-functions are \cite{lm}
$dY_i^{ab}/dt=U_i^{ab}$, $dU_i^{ab}/dt=0$, which are the Galilean evolution
equations for the D-particles if we identify the time $t=\ell_s\log\Lambda$
with the worldsheet scale.

The natural geometry on the moduli space $\cal M$ of deformed conformal field
theories described above is given by the Zamolodchikov metric \cite{zam} which
is the two-point function $G_{ab;cd}^{ij}=2\Lambda^2\langle
V_{ab}^i(x;0)V_{cd}^j(x;0)\rangle$ in the full model \eqn{partfatbrane}. Using
the representation \eqn{wilsonloopaux}, after a long perturbative calculation
we find \cite{ms}
\bea
G_{ab;cd}^{ij}&=&\frac{4g_s^2}{\ell_s^2}\left[\delta^{ij}\,I_N\otimes
I_N-\frac{g_s^2}6\left\{I_N\otimes\left(U^iU^j+U^jU^i\right)+
U^i\otimes U^j\right.\right.\nn\\&
&\biggl.\left.+\,U^j\otimes U^i+\left(U^iU^j+U^jU^i\right)\otimes
I_N\right\}\biggr]_{db;ca}+{\cal O}\left(g_s^6\right)
\label{Gfinal}\eea
where $I_N$ is the $N\times N$ identity matrix and we have renormalized $g_s$
to the time-independent coupling $g_s/|\epsilon|\ell_s$. The metric
\eqn{Gfinal} is a complicated function of the D-particle dynamical parameters,
which will be the key to its use in probing short-distance structure on $\cal
M$. One can also perturbatively calculate the canonical momentum $P_{ab}^i$ of
the D-particle system, which, in the Schr\"odinger picture, is given by the
expectation value of $-i\delta/\delta Y_i^{ab}$ with respect to
\eqn{partfatbrane}, i.e. $P_{ab}^i=\langle V_{ab}^i(x;0)\rangle$. Another long
and tedious calculation gives \cite{ms}
\beq
P_{ab}^i=\frac{8g_s^2}{\ell_s}\left[U^i-\frac{g_s^2}6
\left(U_k^2U^i+U_kU^iU^k+U^iU_k^2\right)\right]_{ba}+{\cal O}\left(g_s^6\right)
\label{canmom}\eeq
which coincides with the contravariant velocity
$P_{ab}^i=\ell_sG_{ab;cd}^{ij}\dot Y_j^{cd}$ on $\cal M$. The final quantity we
need is the Zamolodchikov $C$-function, which interpolates on $\cal M$ among
two-dimensional field theories according to the $C$-theorem $\partial{\cal
C}/\partial t=\e^{-{\cal C}t}\,\dot Y_i^{ab}G_{ab;cd}^{ij}\dot Y_j^{cd}$
\cite{zam}. This differential equation can be solved for small velocities to
give the physical target space time coordinate \cite{emn}
\beq
T\equiv\sqrt{\ell_s\,{\cal
C}(t)}~t\simeq2g_st\,\sqrt{\mbox{$\frac1{\ell_s}$}\,{\cal
F}(U)}\,\left(\int_0^td\tau~\e^{2(\tau^2-t^2)g_s^2{\cal
F}(U)/\ell_s^2}\right)^{1/2}
\label{phystime}\eeq
where we have introduced the scalar function
\beq
{\cal F}(U)=\tr\,U_i^2-\frac{g_s^2}3\,\tr\left[2\,U_i^2U_j^2+(U_iU_j)^2
\right]+{\cal O}\left(g_s^4\right)
\label{Fdef}\eeq
The moduli space dynamics can be derived from the Lagrangian
$-\frac{\ell_s}2\dot Y_i^{ab}G_{ab;cd}^{ij}\dot Y_j^{cd}-{\cal C}$ \cite{emn}
which can be shown~\cite{ms} to coincide to leading orders with the non-abelian
Born-Infeld effective action \cite{tseytlin} for the target space D-particle
dynamics.

The quantization of $\cal M$ comes from summing over all worldsheet genera of
the $\sigma$-model \eqn{partfatbrane}. The dominant contributions come from
pinched annulus diagrams of pinching size $\delta\to0$ \cite{emn}.
Symbolically, this approximation leads to a genus expansion of the form
\unitlength=1.00mm
\linethickness{0.4pt}
\beq
\begin{picture}(70.00,10.00)
\LARGE
\put(5.00,2.00){\circle*{100.00}}
\put(8.00,2.00){\circle{3.00}}
\put(15.00,2.00){\makebox(0,0)[l]{$+$}}
\put(28.00,2.00){\circle*{100.00}}
\put(31.00,2.00){\circle{3.00}}
\put(28.00,5.00){\circle{3.00}}
\put(38.00,2.00){\makebox(0,0)[l]{$+$}}
\put(48.00,2.00){\circle{3.00}}
\put(51.00,2.00){\circle*{100.00}}
\put(54.00,2.00){\circle{3.00}}
\put(51.00,5.00){\circle{3.00}}
\put(61.00,2.00){\makebox(0,0)[l]{$+~\dots$}}
\end{picture}
\label{pinch}\eeq
consisting of thin tubes (worldsheet wormholes) attached to $\Sigma$. The
attachment of each tube corresponds to inserting a bilocal pair
$V_{ab}^i(x;s)V_{cd}^j(x;s')$  on $\partial\Sigma$, with interaction strength
$g_s^2$, and computing the string propagator along the thin strips. There are
modular $\log\delta$ divergences that arise, which should be identified with
worldsheet divergences at lower genera \cite{emn}, and hence we
take $\log\delta=2g_s^\eta\log\Lambda=\frac1{\ell_s^2}g_s^\eta\epsilon^{-2}$
(the exponent $\eta\geq0$ is left arbitrary for the moment in order to compare
with other results later on). The effects of the dilute gas of wormholes
\eqn{pinch} on $\Sigma$ are to exponentiate the bilocal operator leading to a
change in the action in (\ref{partfatbrane}, \ref{wilsonloopaux}). This
contribution can be cast into the form of a local action by rewriting it as a
functional Gaussian integral over wormhole parameters $\rho_i^{ab}$, and we
arrive finally at \cite{ms}
\beq
\sum_{\rm genera}Z_N[Y]\simeq\left\langle\int_{\cal
M}D\rho~\e^{-\rho_i^{ab}G_{ab;cd}^{ij}\rho_j^{cd}/2|\epsilon|^2\ell_s^2
g_s^2\log\delta}~W[\partial\Sigma;Y+\rho]\right\rangle_0
\label{genusexp}\eeq
{}From \eqn{genusexp} we see that the effect of the resummation over pinched
genera is to induce quantum fluctuations of the solitonic background, giving a
statistical Gaussian spread to the D-particle couplings. Note that the width of
the Gaussian distribution in \eqn{genusexp}, which we identify as the
wavefunction of the system of D-particles \cite{emn}, is time independent
and represents not the spread in time of a wavepacket on $\cal M$, but rather
the true quantum fluctuations of the D-particle coordinates.

The corresponding spatial uncertainties can be found by diagonalizing the
Zamolodchikov metric \eqn{Gfinal}. We apply a Born-Oppenheimer approximation to
the D-particle interactions, which is valid for small velocities
\cite{liyoneya}, to separate the diagonal D-particle coordinates from the
off-diagonal parts of the Higgs fields representing the short open string
excitations connecting them. This is achieved via a time independent gauge
transformation $Y^i=\Omega\,{\rm diag}(y^i_1,\dots,y^i_N)\,\Omega^{-1}$, where
$\Omega\in U(N)$ describes the fundamental strings and the eigenvalues
$y_a^i\in\real$ represent the positions of the D-particles which move at
velocities $u_a^i=\dot y_a^i$. This diagonalizes \eqn{Gfinal} in its colour
indices. To diagonalize the resulting tensor in its spacetime indices
$i,j=1,\dots,9$ we consider the case where, for simplicity, two given
D-particles $a,b$ move orthogonally to each other, as depicted in fig. 1, with
the same speed $|u_a|=u$. For $a=b$, there is only one non-zero eigenvalue
$\lambda_{aa}^1(u)=6u^2$, owing to the fact that a single D-particle on its own
has only open string excitations starting and ending on itself. Upon rotation
to the one-dimensional frame spanned by the velocity vector $u_a^i/u$, the
spacetime diagonalizing matrix $O_{aa}(u)$ is the $9\times9$ identity matrix.
We shall refer to this frame as the `string frame', because it represents the
coordinate system relative to the open string excitations of the D-particles.
For $a\neq b$, there are two non-zero eigenvalues
$\lambda_{ab}^{\pm}(u)=(2\pm\sqrt5)u^2$, and the dimension of the string frame
increases by one since in this case the string stretches between two different
D-particles. We parametrize the plane spanned by $u_a$ and $u_b$ as
$u_a^i=u\delta^{i,1}$ and $u_b^i=u\delta^{i,2}$, and then the orthogonal
diagonalizing matrix $O_{ab}(u)$ has a block diagonal form consisting of the
$2\times2$ block matrix $\frac1{\sqrt6}{\scriptstyle
 \addtolength{\arraycolsep}{-.5\arraycolsep}
 \renewcommand{\arraystretch}{0.5}
 \left( \begin{array}{cc}
 \scriptstyle \sqrt5 & \scriptstyle -1 \\
 \scriptstyle 1  & \scriptstyle \sqrt5 \end{array} \scriptstyle\right)}$ and
the $7\times7$ identity matrix.

The coordinate transformation
${\tilde Y}_{ab}^i=O_{ab}^{ji}[\Omega^{*-1}Y_j\Omega]_{ba}\equiv
O^{ji}_{ab}X^j_{ab} $, where the collective D-particle coordinates $X^i(Y)$
contain information about the open string excitations, diagonalizes the
bilinear form in \eqn{genusexp} and yields the statistical variances
$(\Delta{\tilde Y}^i_{ab})(\Delta{\tilde Y}^i_{ab})^\dagger=\ell_s^2
g_s^\eta(1+\frac{g_s^2}{48\pi^3}\lambda_{ab}^i(u))+{\cal O}(g_s^4)$. For $a=b$
we therefore arrive at the uncertainties
\be
\left|\Delta
X^i_{aa}\right|=\ell_sg_s^{\eta/2}\left(1+\frac{g_s^2}{8\pi^3}\,u^2\,
\delta^{i,1}\right)+{\cal O}\left(g_s^4\right)\geq\ell_s\,g_s^{\eta/2}
\label{minimum}\ee
for the individual D-particle coordinates. For $\eta=0$ the minimal length in
\eqn{minimum} coincides with the standard string smearing \cite{ven} while for
$\eta=\frac23$ it matches $\ell_{\rm P}$ which arises from the kinematical
properties of D-particles \cite{liyoneya}. A choice of $\eta\neq0$ is more
natural since the modular strip divergences should be small for weakly
interacting strings. Note that the uncertainty \eqn{minimum} is always larger
in the string frame, representing the additional smearing that occurs from the
open string excitations on the D-particles. Outside of this frame we obtain
exactly the standard stringy smearings directly from the worldsheet formalism,
without the need of postulating an auxilliary uncertainty relation as is done
in \cite{ven,liyoneya}.

The coordinate uncertainties for $a\ne b$ are responsible for the emergence of
a true noncommutative quantum spacetime and represent the genuine non-abelian
characteristics of the D-particle dynamics. In this case we find the same
velocity independent spreads as in \eqn{minimum} for the coordinates in the
space orthogonal to the two-dimensional string frame. In the string frame we
may assume, by symmetry, that $\Delta X_{ab}^1\sim\Delta X_{ab}^2$, and then we
arrive at a system of two linear equations in two unknowns. Solving them
simultaneously gives a minimum length uncertainty $\Delta X_{ab}^1$ of the same
form as \eqn{minimum} and the additional correlator
\beq
{\rm Re}\left\langle\!\left\langle
X_{ab}^1\,X_{ab}^2\right\rangle\!\right\rangle_{\rm
conn}^{(\rho)}=\frac{\ell_s^2}{16\pi^3}\,g_s^{2+\eta}\,u^2~~~~~~,~~~~~~a\neq b
\label{x1x2}\eeq
where $\langle\langle\,\cdot\,\rangle\rangle_{\rm conn}^{(\rho)}$ denotes the
connected statistical correlation function with respect to the wormhole
probability distribution in (\ref{genusexp}). When the D-particles are in
motion, we see from \eqn{x1x2} that the associated open string position
operators are not independent random variables and have a non-trivial quantum
mechanical correlation. This is a new form of spacetime quantum uncertainty
relations between different {\it spatial} directions of the target space. For
$\eta=\frac23$, the right-hand side of \eqn{x1x2} can be written in terms of
$\ell_{\rm P}^2$ when we identify the kinetic energy of the D-particles with
$g_s^2u^2$, as follows from \eqn{canmom}. The energy dependence of
\eqn{minimum} and \eqn{x1x2} is a quantum decoherence effect which can be
understood from a generalization of the Heisenberg microscope whereby we
scatter a low-energy probe, represented by a closed string state, off the
D-particle configuration (see fig. 1). As the closed string state hits a
D-particle, it splits into two open string states, represented by the recoil of
the particle upon impact with the detector, which absorb energy from the
scattering. For an isolated D-particle, these open string excitations have
their ends attached to the same point. For two D-particles the ends of the open
string can attach to different points. Since the recoil of the constituent
D-particles causes the bound state configuration to recoil as well, the
interactions mediated by the open strings cause a non-trivial correlation
between different coordinate degrees of freedom stretched between the two
particles. Only when there is no recoil ($u=0$) can one measure independently
the positions of the two D-particles.

We can derive space-time uncertainty principles by exploiting the canonical
structure on $\cal M$ represented by \eqn{canmom}. In the present context,
whereby quantum mechanical averages are identified with vacuum expectation
values of the $\sigma$-model on $\Sigma$, the uncertainty $(\Delta P_{ab}^i)^2$
in the momentum can be computed as the difference between \eqn{Gfinal} for
$a=c,b=d,i=j$ and the square of \eqn{canmom}. For weakly coupled strings this
yields $(\Delta P_{ab}^i)^2=\frac{4g_s^2}{\ell_s^2}\delta_{ab}+{\cal
O}(g_s^4)$, and using the minimal length uncertainties computed above we find
the canonical uncertainty relation $\Delta Y_i^{ab}\Delta
P^j_{cd}\geq2g_s^{1+\eta/2}\delta_i^j\delta_c^a\delta_d^b$. Upon summation over
genera the target space time coordinate \eqn{phystime} becomes a quantum
operator \cite{aemn}, unlike the situation in conventional quantum mechanics.
Within the present Born-Oppenheimer approximation, we can expand the function
\eqn{Fdef} as a power series in $|U_{ab}|/u\ll1$, $a\neq b$. Then using the
leading Heisenberg commutation relations between $Y$ and $P\sim8g_s^2U/\ell_s$,
we arrive at the space-time commutation relation
\beq
\left[Y_i^{ab},T\right]\simeq2i\,\ell_s^2\,g_s^{\eta/2}
\left(\delta^{ab}+\left(1-\delta^{ab}\right)\frac{U_i^{ba}}{2u}\right)
\label{na}\eeq
to leading order in $g_s$ (or equivalently in the off-diagonal velocity
expansion). From \eqn{na} we infer the space-time uncertainty relation $\Delta
Y_i^{aa}\Delta T\geq\ell_s^2g_s^{\eta/2}$ for the D-particle coordinates, which
for $\eta=0$ yields the standard lower bound \eqn{tyuncert} which is
independent of $g_s$, as argued in \cite{liyoneya} from a basic string
theoretic point of view. But then the minimal distance \eqn{minimum} doesn't
probe scales down to $\ell_{\rm P}$. This fact can be understood by noting that
$T$ is not the same as the longitudinal worldline coordinate of a D-particle,
as is assumed in \eqn{tyuncert}, but is rather a collective time coordinate of
the D-particle system which is induced by all of the interactions among the
particles. However, we can adjust the uncertainty relations to match the
dynamical properties of 11-dimensional supergravity by multiplying the
definition (\ref{phystime}) by a factor $g_s^{-\eta/2}$, which implies that
with weak string interactions the target space propagation time for the
D-particles is very long.

To see the effects of the string interactions between D-particles, we again use
the canonical smearing between $Y^{ab}_j$ and $P_{ab}^i$ for $a\ne b$ in
\eqn{na} to arrive at a triple uncertainty relation
\be
\left(\Delta Y_i^{ab}\right)^2\Delta
T\geq\frac{\ell_s^3\,g_s^{\eta/2-1}}{16u}~~~~~~,~~~~~~a\neq b
\label{triple}\ee
The uncertainty principle (\ref{triple}) implies that the system of
D-particles, through their open string interactions, can probe distances much
smaller than the characteristic distance scale in \eqn{triple}, which for
$\eta=\frac23$ is $\ell_{\rm P}\ell_s^2$, provided that their kinetic energies
are large enough. Of course, in the fully relativistic case the existence of a
limiting velocity $u<1$ implies a lower bound on \eqn{triple}. With the minimum
spatial extensions obtained above, this bound yields the characteristic
temporal length $\Delta T\geq g_s^{-1/3}\ell_s$ for D-particles
\cite{liyoneya}. Triple uncertainty relations involving only $\ell_{\rm P}^3$
have been suggested in \cite{liyoneya} based on the holographic principle of
M-theory.

The kinetic energy dependence of \eqn{x1x2} and (\ref{triple}) is the main
distinguishing feature between D-particle dynamics and ordinary quantum
mechanics, since it implies a bound on the measurability of lengths that
depends entirely on the energy content of the system. Such uncertainties, where
the bound on the accuracy with which one can measure a quantity depends on its
size, are characteristic features of decoherence in certain approaches to
quantum gravity \cite{aemn}. In the present case the quantum coordinate
fluctuations, due to the open string excitations between D-particles, can lead
to quantum decoherence for a low energy observer who cannot detect such recoil
fluctuations in the sub-Planckian spacetime structure \cite{emn}. This approach
to short-distance physics using non-abelian D-particle dynamics therefore seems
to have naturally encoded within it features of quantum gravity. These foamy
properties of the noncommutative structure of the D-particle spacetime might
require a reformulation of the phenomenological analyses of length measurements
as probes of quantum gravity. More details about these constructions will
appear in a forthcoming paper \cite{ms}.

\end{document}